%% file: decorr_amm.tex
\documentclass[
twocolumn, 
superscriptaddress,
nofootinbib,
amsmath,amssymb,
pre,
]{revtex4-1}

\usepackage{graphicx}
\usepackage{subfigure}
\usepackage{color}
\usepackage{xspace}
\usepackage{amsmath,amssymb,bm}
\usepackage{cases}
\usepackage{tikz}
  \usetikzlibrary{calc}
  \usetikzlibrary{arrows}
  \usetikzlibrary{decorations.markings}
\usepackage{afterpage}
\usepackage{ulem}
\usepackage{hyperref}

\expandafter\ifx\csname package@font\endcsname\relax\else
  \expandafter\expandafter
  \expandafter\usepackage
  \expandafter\expandafter
  \expandafter{\csname package@font\endcsname}%
\fi

\input{macrosGunnar}

\usepackage{dsfont}

\newcommand{\canetset}[1]{{\mathchoice {\hbox{$\sf\textstyle #1\kern-0.4em #1$}}
{\hbox{$\sf\textstyle #1\kern-0.4em #1$}}
{\hbox{$\sf\scriptstyle #1\kern-0.3em #1$}}
{\hbox{$\sf\scriptscriptstyle #1\kern-0.2em #1$}}}}

\newcommand{\profile}{\rho_\parallel}

\newcommand{\Frefs}[1]{Figs.~\ref{fig:#1}}

\newcommand{\bibconferencename}[1]{\textit{#1}}

\begin{document}

\title{Correlation, crossover and broken scaling in the Abelian Manna Model}

\author{Letian Chen}
\affiliation{Department of Mathematics, Imperial College London, United Kingdom}
% \email{letian.chen16@imperial.ac.uk}

\author{Hoai Nguyen Huynh}
\thanks{HNH is currently not affiliated with Imperial College London.}
\affiliation{Department of Mathematics, Imperial College London, United Kingdom}
\affiliation{Institute of High Performance Computing, Agency for Science, Technology and Research, Singapore}

% \altaffiliation[On leave from: ]{Institute of High Performance Computing, Agency for Science, Technology and Research, Singapore}
%[leave on:]
% \email{n.huynh@imperial.ac.uk}
% \homepage{https://sites.google.com/site/nelive}

\author{Gunnar Pruessner}
\email{g.pruessner@imperial.ac.uk}
\affiliation{Department of Mathematics, Imperial College London, United Kingdom}
\homepage{http://www.ma.ic.ac.uk/~pruess}

% \author{Letian Chen}
% \affiliation{Department of Mathematics, Imperial College London, United Kingdom}
%  \email{letian.chen16@imperial.ac.uk}

% \author{Hoai Nguyen Huynh}
% \affiliation{Department of Mathematics, Imperial College London, United Kingdom}
% \altaffiliation[On leave from: ]{Institute of High Performance Computing, Agency for Science, Technology and Research, Singapore}
%  \email{n.huynh@imperial.ac.uk}
% \homepage{https://sites.google.com/site/nelive}

% \author{Gunnar Pruessner}
% \affiliation{Department of Mathematics, Imperial College London, United Kingdom}
% \email{g.pruessner@imperial.ac.uk}
% \homepage{http://www.ma.ic.ac.uk/~pruess}

\begin{abstract}
%The role of correlation and symmetry in self-organised critical phenomena is
%investigated by studying a proptypical model. The model is studied on lattices
%in one and two dimensions. The local correlations are destroyed by re-arranging
%particles on the lattice between avalanches. It is found that as long as spatial
%symmetry is preserved in one dimension, the local correlations do not change the
%dynamical behaviour of the system. However, simple preservation of spatial
%symmetry is not enough in two dimensions, in which the critical behaviour
%becomes that of a trivial type.
The role of correlations in self-organised critical (SOC) phenomena is
investigated by studying the Abelian Manna Model (AMM) in two dimensions. 
Local correlations of the debris left behind after avalanches are destroyed by re-arranging particles on the lattice between avalanches, without changing the one-point particle density. It is found
that the spatial correlations are not relevant to small avalanches, while changing the scaling of the large (system-wide) ones, yielding a crossover in the model's scaling behaviour. This crossover breaks the simple scaling observed in normal SOC.
% producing a ``multi-scaling'' behaviour, which can be captured by introducing a so-called multi-scaling \emph{ansatz}. This \emph{ansatz} is well validated by a data collapse.
% \gcomment{The message should be: Yes, there are some mild correlations, but
% we want to know whether they have any bearing in the scaling.
% I wonder whether we would be able to get a reasonable $P(s)$ starting from
% a single config (repeatedly).} 
% \lecomment{Is multi-scaling \emph{ansatz} one of the main results? }
\end{abstract}

% \keywords{Self-organised criticality,
% Universality,
% Correlation functions,
% Finite-size scaling}

\maketitle

%\tableofcontents

\section{\label{sec:level1}INTRODUCTION}

Spatial correlations are at the core of self-organised critical (SOC) phenomena, because part of the narrative of SOC is the spreading of long-range spatial correlations as systems evolve towards the critical point \cite{WatkinsETAL:2016}.
Correlations between (but also within) avalanches are thought to be mediated by ``substrate particles'' \cite{DickmanVespignaniZapperi:1998}, \ie, the immobile particles left behind by avalanches. These are replenished by the repeated external driving and form the fuel and backdrop of the next avalanche. 
Despite the great conceptual importance to SOC, spatial correlations are rarely studied directly, as they are notoriously noisy and difficult to analyse \cite{Pruessner:2012:Book}.
In the Manna Model \cite{Manna:1991a,Dhar:1999a}, Hexner and Levine  \cite{HexnerLevine:2015} identified hyperuniformity in the substrate particle density and Basu \etal \ \cite{BasuETAL:2012} observed ``natural long-range correlations in the background”. Using the fact that natural critical states in the Oslo rice pile model are hyperuniform, Grassberger \etal \ \cite{GrassbergerDharMohanty:2016} were able to perform vastly improved simulations. Recently, one of us \cite{WillisPruessner} found the algebraic auto-correlations of immobile substrate particles in the Manna Model to display hyperuniformity with comparatively small amplitude when measured directly. As the density of immobile particles is bounded, the amplitude of their auto-correlations cannot increase with system size. 
If coarse graining was to be applied, any such correlations can therefore be made arbitrarily small at any non-vanishing distance, in contrast to the invariance under rescaling normally observed in critical phenomena and confirmed for the \textit{activity} in the Manna Model \cite{WillisPruessner}.

Similarly, the density of immobile particles as a function of position
on the lattice (introduced as the ``density profile'' $\rho(x,y)$ below) rises quickly away from the boundaries, remains (almost) constant in the bulk and then drops sharply again at the other boundary, Fig.~1 in \cite{WillisPruessner}. 

The aim of the present work is to determine \emph{whether the correlations of substrate particles have a bearing on the critical state in the sense
of being a prerequisite of the critical behaviour in the Manna Model}, rather than being a result of it. Are these correlations weak but important, or weak and irrelevant? 

One may attempt to address this question by redistributing particles uniformly in the stationary state. In the Manna Model this was investigated by Stapleton (Chapter 5 in \cite {Stapleton:2007}) in one dimension, resulting in scaling compatible with the Manna universality class. One of us repeated some of these numerical experiments on much bigger lattices \cite[p.~172, 173]{Pruessner:2012:Book} and found deviating exponents, albeit non-trivial ones. However, by redistributing, not only the correlations are destroyed, but also the density profile is evened out. Field-theoretically, the density profile is expected to be more relevant than the two-point correlation function \cite{LeeCardy:1995} not least as its shape has a direct impact on particle transport towards boundaries \cite{PaczuskiBassler:2000v2} and thus on the avalanche size at the stationary state. We will therefore test the relevance of the correlations by removing or diluting any auto-correlations that may have developed over the course of an avalanche without, however, destroying the density profile. We will then probe whether the critical behaviour is effected.

At this stage, we deliberately allow for ambiguity in the notion of ``bearing on the critical behaviour'' --- a possible outcome may be that after tempering with the substrate particle distribution the Manna Model remains critical, yet displays scaling behaviour different from that of the Manna universality class. Another outcome may be loss of all (non-trivial) critical behaviour whatsoever.

In the following, we will first introduce the Manna Model and the relevant
observables in detail in \sref{AMM}, then show different setups used to destroy the correlations in the substrate in \sref{initialisation}. The results are presented in \sref{Results} and \Aref{appendixa}. In \sref{discussion}, we will discuss the different exponents we find and the role of correlations. \sref{conclusion} contains a summary of our results, and some concluding remarks.

\section{MODEL}
\Slabel{Model}
\subsection{Abelian Manna Model}
\Slabel{AMM}

The Abelian variant of the Manna Model \cite{Manna:1991a,Dhar:1999a} used in the present work is defined as follows: Consider a two-dimensional square lattice with linear size $L$. Each site, indexed by $(x,y)\in\{1,2,\ldots,L\}^{2}$, can hold a non-negative integer number of particles, denoted by $z(x,y)$. A site $(x,y)$ is called \textit{stable} if the number $z(x,y)$ of particles it contains does not exceed a threshold of one, \ie, $z(x,y) \leq 1$. When all sites are stable, the system is in a \textit{quiescent} state. On the other hand, a site $(x,y)$ is called \textit{unstable} or \textit{active} if $z(x,y) > 1$. Any active site relaxes by toppling, whereby two of its particles are randomly and independently redistributed to its nearest neighbouring sites. The space outside the lattice may be thought of as a neighbouring site that never topples itself: If a toppling takes place at a site next to an open boundary, up to two particles can get lost from the system altogether, if they are being placed beyond the boundary by the toppling. These particles are said to have been \textit{dissipated}.  

Initially, there is no particle on the lattice. The system is loaded by the external driving mechanism, which adds one particle to a randomly chosen site provided the system is quiescent. If the chosen site already contains one particle, a toppling will be triggered. This toppling may cause the neighbour of the driven site to become active, leading to further topplings. No more toppling occur when all sites are stable, \ie, the system is quiescent. At this point, a new driving event occurs at a random site, independent of previously driven sites, potentially triggering a new series of topplings. The series of topplings between one external driving and the system returning to a quiescent state is called an avalanche.
The external driving is ``slow'' because it occurs only once the avalanche has stopped, creating a separation of the time scales of driving and relaxation. This separation is thought to be a key feature of SOC \cite{Pruessner:2012:Book,WatkinsETAL:2016}. The particles left behind after an avalanche, the ``debris'', will be referred to as the \textit{substrate} within which the next avalanche takes place. As avalanches transport particles to the boundaries \cite{PaczuskiBassler:2000v2} eventually resulting in their dissipation, the expected number of topplings following a driving is finite, because any avalanche sooner or later runs out of ``fuel'', resulting in correlations between successive avalanches \cite{GarciaMillanPruessner:2018}.

The total number of topplings that occur during an avalanche defines the avalanche size, $s$. The number of distinct sites receiving a particle during an avalanche is called the avalanche area $a$. If no toppling occurs after the driving, the avalanche size is $0$, and the avalanche area is $1$. Multiple sites may be active simultaneously during the toppling sequence. The Abelian nature \cite{Dhar:1990a} of the model refers to the fact that the final stable state of the system is invariant under changes of the updating order of unstable sites \cite{Pruessner:2012:Book}. The microscopic time increases by $1/k$ if any of the $k$ simultaneously active site topples. The microscopic time between an external driving and the system returning to a quiescent state is called the avalanche duration, $T$. If no toppling takes place after the driving, we define $T = 0$. 
Additionally, we measure the instantaneous particle density $\zeta$ in the quiescent state, \ie, after avalanches have ceased, given by
\begin{equation}
\zeta = \frac{1}{L^2} \sum_{x,y} z(x,y)\ \text{,}
\end{equation}
which fluctuates around a mean value in the stationary state when the influx of particles balances the outflux.

\begin{figure}
% \centering
\includegraphics[width=1\columnwidth]{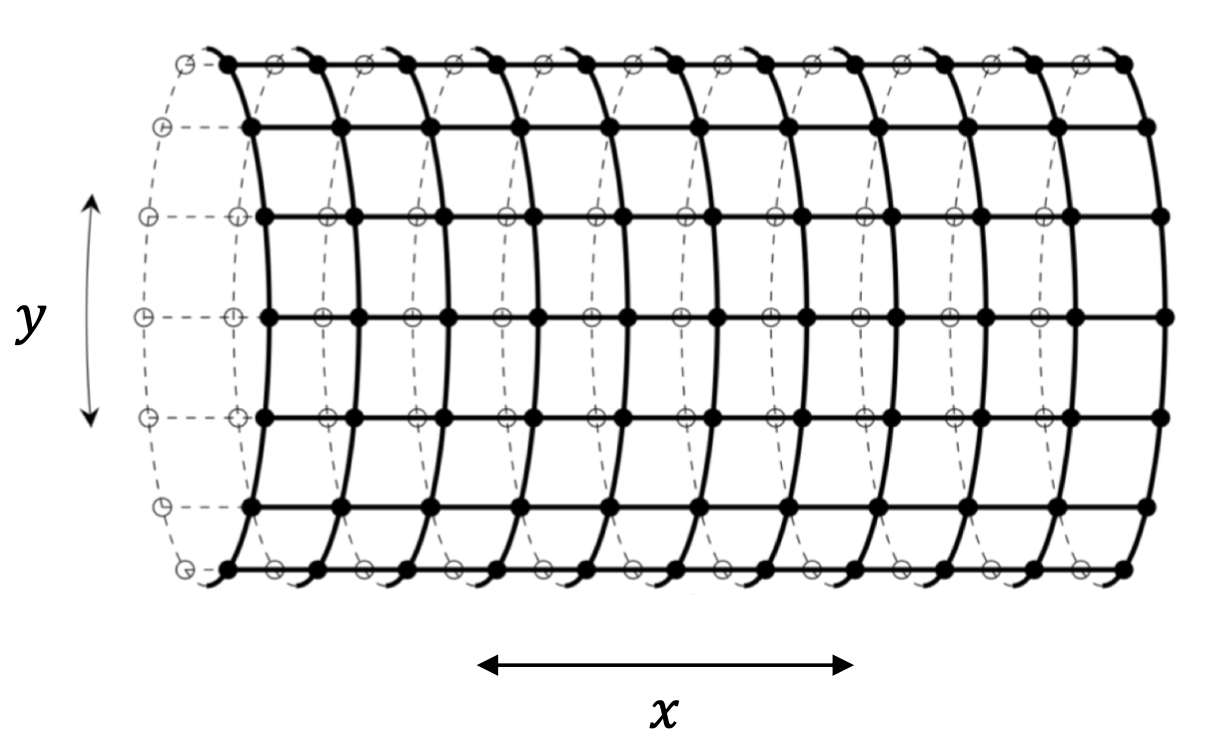}
\caption{\label{2D_periodic} Square lattice with open boundary conditions in the $x-$direction and periodic boundary conditions in the $y-$direction, which resembles a cylinder.}
\end{figure}

\subsection{Boundary conditions and re-initialisation}
\Slabel{initialisation}

In this work, we investigate the Abelian Manna Model (AMM) on two-dimensional square lattices, with periodic boundary conditions in the $x-$direction and open boundary conditions in the $y-$direction, Fig.~\ref{2D_periodic}. We will call this setup the \textit{periodic model}. Apart from the periodic boundaries in one direction, the \textit{periodic model} is identical to the \textit{original AMM}, which has open boundaries in all dimensions, and they belong to the same universality class 
%\cite{PruessnerHuynh:2012}. 
\cite[Tab.~9.3 in][]{Huynh:2013}.
The periodic boundary condition identifies site $(x,y)$ with site $(x,y+L)$. A set of sites with the same $x$ will be called a \textit{ring}. The presence of open boundaries in the $x-$direction introduces dissipative transport (\cite{Grinstein:1995}, Chapter 9.2 in Ref.~\cite{Pruessner:2012:Book}). Particles added to the system by external driving may leave the lattice through the open boundary after a series of topplings. In fact, particles in the system perform random walks. The average avalanche size $\ave{s}$ can be determined in closed form by mapping it to the expected number of steps of a random walker before escaping from the lattice. In this context and the following, $\ave{\cdot}$ represents the ensemble average of an observable in the stationary state. For a $d$-dimensional square lattice with open boundary conditions in one direction and periodic boundary conditions in other $d-1$ directions, the expected avalanche size is \cite{Pruessner_aves_Manna:2013}
\begin{equation}
\elabel{first_moment_size}
\langle s\rangle = d\frac{(L+1)(L+2)}{12}\ \text{,}
\end{equation}
based on the residence time of a random walk. An avalanche leaves behind an array of particles in a quiescent system, that may be thought of as the substrate upon which the next avalanche will be running. Avalanche therefore leaves an imprint on the lattice, such that the substrate particles' spatial-temporal features are correlated. The expected occupation number $\rho(x,y)$ at site $(x,y)$ in the quiescent state, referred to as the \emph{density profile}, is defined as $\rho(x,y)=\ave{z(x,y)}$ with sum rule $\sum_{x,y}\rho(x,y)=L^2\ave{\zeta}$. The density profile is translational invariant along the $y-$direction due to the periodic boundary conditions, \ie
\begin{equation}
\label{periodic_condition}
    \forall_{a\in\{1,\ldots,L\}} \rho(x,a)=\profile(x) \ .
	% \forall_{a\in\Zset} \rho(x,y+a)=\rho(x,y)=\profile(x)
\end{equation}
This property does not apply to the $x-$direction, \ie, generally $\rho(x+a,y)\ne\rho(x,y)$, since translational invariance in the open direction is broken due to the open boundary conditions. In fact, the density profile in the $x-$direction shows a dip towards the open boundaries  \cite{WillisPruessner}, because of the transport towards the open ends \cite{PaczuskiBassler:2000v2}. We expect that the density profile has a bearing on the behaviour of the AMM possibly even affecting its critical behaviour, as the density profile immediately affects the particle transport.

We further introduce the connected correlation function
\begin{equation}
\begin{split}
\label{def_corr}
C(x_1,y_1; x_2,y_2) = &\ave{z(x_1,y_1)z(x_2,y_2)} - \\
                     & \ave{z(x_1,y_1)}\ave{z(x_2,y_2)} ,
\end{split}
\end{equation}
which has been measured in \cite{WillisPruessner} and was found to be surprisingly weak (see Fig.~2 in \cite{WillisPruessner}). The present work centres around the question whether the weak correlations matter for the critical behavior.
To test the effect of correlations of substrate particles, we deliberately destroy them by re-arranging particles between avalanches. If correlations are weak yet important, the critical properties of the Manna Model with the correlations destroyed should differ from those in the \textit{periodic model}. If correlations are weak and irrelevant, the critical behaviour should remain unchanged. To destroy correlations without unduly changing the density profile and thus tampering with the (critical) behaviours in a ``trivial''  manner, we introduce four different operations that destroy the correlations of substrate particles while preserving the instantaneous density profile of the substrate. The operation of choice is applied after every non-vanishing avalanche, \ie, one that has size $s > 0$. This is equivalent to applying the operation after every avalanche, even those with non-vanishing size, because only a non-vanishing avalanche can recreate the correlation that are destroyed by the operation. Performing this operation even after vanishing avalanches does not change the substrate particles' statistics.

\textbf{Twisting:} One method of re-arranging particles is rotating or twisting periodically closed rings of particles in the direction of the periodic boundary condition, by a random angle independently chosen for every ring. To this end, we draw random integer variables $r_x$ with $0\le r_x<L$ for each $x\in\{1,2,\ldots,L\}$ and re-initialize the lattice with the configuration $z'(x,y)=z(x,y+r_x)$, subject to the periodic boundary conditions, before the next driving. In other words, after each non-vanishing avalanche, we twist every periodic ring of the lattice independently. Particles within the same ring maintain their relative positions, while particles in neighbouring rings do not. This operation destroys correlations in the open $x-$direction so that \textit{after twisting}, in a large enough system and assuming finite correlation length, the correlation function asymptotically factorizes, $C(x_1,y_1; x_2,y_2) \propto \delta_{x_1,x_2}$. If correlations in the substrate are present but irrelevant, then the twisting will not result in a critical behaviour different from the \textit{periodic model}.

\begin{figure}
% \centering
\includegraphics[width=0.6\columnwidth]{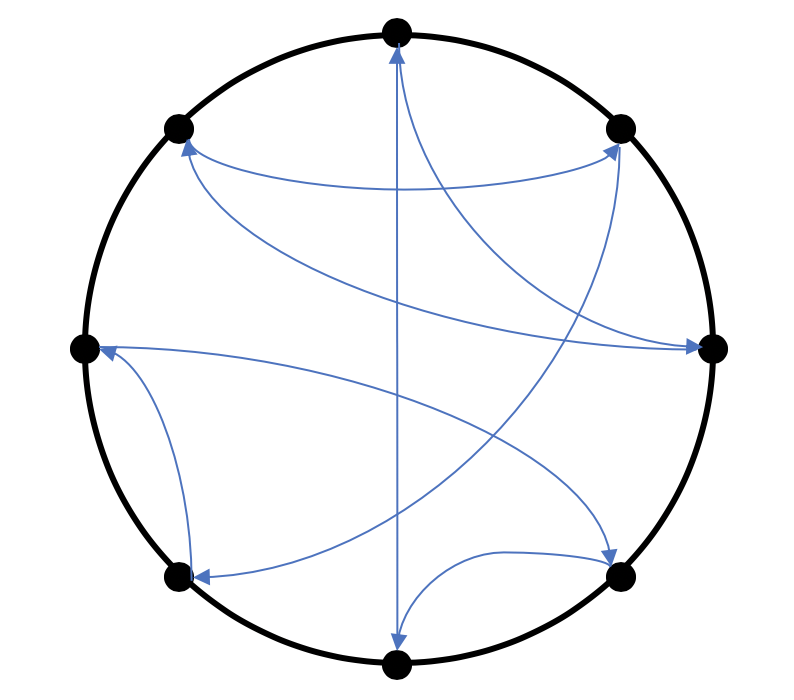}
\caption{\label{swap} Side view of the \textit{periodic AMM} with correlations along the $y-$direction destroyed by \textit{swapping}. The arrows indicate the movement of slabs on the lattice by \textit{swapping}.}
\end{figure}

\textbf{Swapping:} The second method destroys correlations in the $y-$direction by randomly swapping slabs cut along the $x-$direction. As illustrated in Fig.~\ref{swap}, we draw new positions $y'(y)$ for each $y\in\{1,2,\ldots,L\}$ such that $\{y'(1),y'(2),\ldots,y'(L)\}$ is a random permutation of $\{1,2,\ldots,L\}$ and re-initialize the lattice with the configuration $z'(x,y)=z(x,y'(y))$ after each non-vanishing avalanche. Again, particle numbers within a ring remain unchanged. Measuring correlations \textit{after swapping} produces a correlation function that asymptotically factorizes in $y$, $C(x_1,y_1; x_2,y_2) \propto \delta_{y_1,y_2}$, provided the system is large enough and the correlation length is finite.

\begin{figure}
% \centering
\includegraphics[width=1\columnwidth]{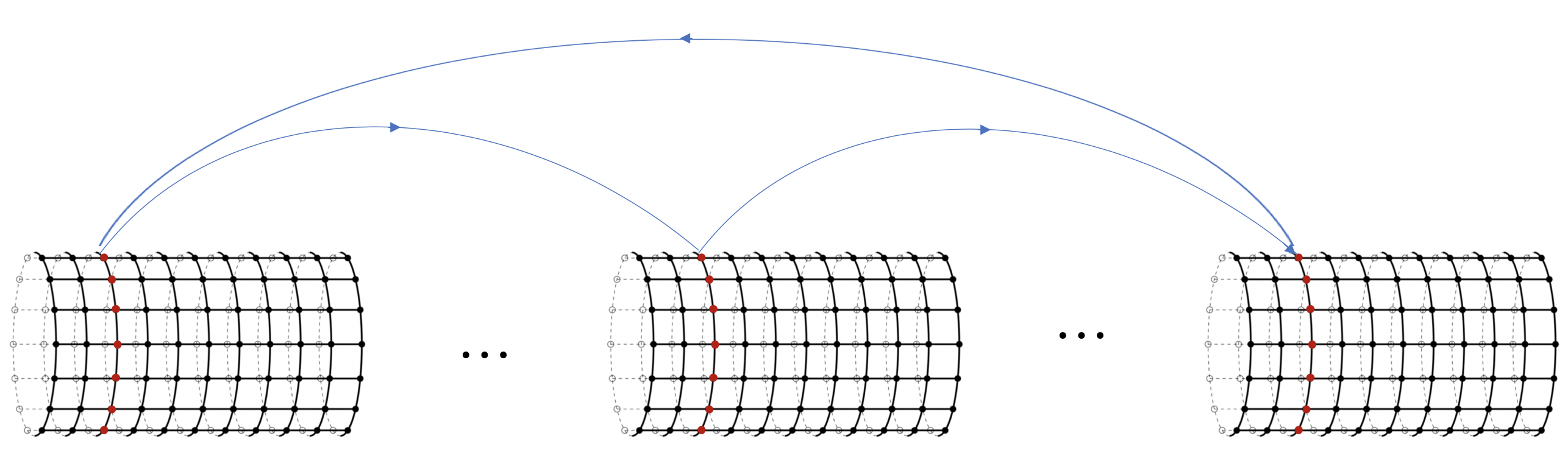}
\caption{\label{shuffling} The schematic of the \textit{periodic AMM} with correlations along the $x-$direction destroyed by \textit{shuffling}. The arrows indicate the possible movement of rings among different systems by \textit{shuffling}.
}
\end{figure}

\textbf{Shuffling:} The third method, which we call \textit{shuffling}, destroys correlations in the $x-$direction by re-arranging rings between different systems within an \emph{ensemble}.  Here, \emph{ensemble} refers to a set of $n$ independent \textit{periodic AMMs}. In this work, we chose $n=10$. The number $z_{i}(x,y)$ of particles on each site $(x,y)$ now also depends on the index $i \in \{1,2,\dots,n\}$ of the systems. Before \textit{shuffling}, every system in the \emph{ensemble} is charged simultaneously and independently, followed by possible avalanches in each system. After every system reaches their quiescent state, for every given $x$, rings of constant $x$ are permutated between the $n$ systems, \ie, $z_{i}(x,y)$ takes the value of $z_{j}(x,y)$ for all $y$, where $j$ is determined from a random permutation of $\{1,2,\dots,n\}$ for each $x$ independently, as illustrated in Fig.~\ref{shuffling}. \textit{After shuffling}, the correlations in the $x-$direction in every system are destroyed.

\textbf{Grinding:} The fourth method destroys correlations in both the $x$ and $y-$directions simultaneously by combining \textit{twisting} and \textit{swapping}. By \textit{twisting}, sites lose the correlations to their neighbors in the $x-$direction. By \textit{swapping}, they lose the correlations to their neighbors in the $y-$direction. As a result, \textit{after grinding} the correlations in both the $x$ and $y$ directions are destroyed.

\section{Results}
\Slabel{Results}

We have performed extensive numerical simulations for the Abelian Manna Model (AMM) in two dimensions with the four operations discussed above. Its scaling behaviour have been characterised by means of \emph{moment analysis} \cite{TebaldiDeMenechStella:1999v1,Luebeck:2000}, and further validated by data collapses of the \emph{probability density functions} (PDF) (Chapter 7.4 in Ref.~\cite{Pruessner:2012:Book}). Large systems with a wide range of sizes have been implemented in this work. For the \textit{AMM with twisting}, the linear sizes considered were $L \in \{257, 513, 1025, 2049, 4097,8193\}$. For the \textit{AMM with swapping}, \textit{shuffling} and \textit{grinding}, due to the higher demands on CPU-time needed to perform the re-arrangement of particles, the largest system size considered was smaller, but not less than $L=2049$. We will present and discuss results for the \textit{AMM with twisting} in details in the following. Consistent results were found for systems with \textit{swapping}, \textit{shuffling} and \textit{grinding}, presented in \Aref{appendixa}.

\subsection{Moment Analysis}
\Slabel{moment_analysis}

Moment analysis (Chapter 7.3 in Ref.~\cite{Pruessner:2012:Book}) is a standard technique to extract critical exponents for a quantity $x$ whose probability density distribution (PDF) is assumed to follow the \textit{simple} finite-size \textit{scaling ansatz} for $x\gg x_0$
\begin{equation}
\elabel{simple_scaling_ansatz}
\mathcal{P}_{x}(x;L)=a_{x}x^{-\tau_{x}}\mathcal{G}_{x}(x/x_{c}(L)), \text{with}\  x_{c}(L)=b_{x}L^{D_x},
\end{equation}
 where $L$ is the system size and $x_0$ is the fixed lower cutoff, below which the PDF follows a non-universal function. The two amplitudes $a_x$ and $b_x$ are non-universal metric factors. Here $x_c$ is the upper cutoff above which the PDF deviates significantly from a pure power-law distribution $x^{-\tau_{x}}$ and is dominated by the so-called scaling function $\mathcal{G}_{x}$. Moment analysis is based on the moments
\begin{equation}
\elabel{momentdef}
\ave{x^n}_L = \int_{0}^{\infty}\mathcal{P}_{x}(x;L)x^{n}dx\ \text{,}
\end{equation}
 which according to \Eref{simple_scaling_ansatz} behave like \cite{DeMenechStellaTebaldi:1998,ChristensenETAL:2008}
\begin{equation}
\elabel{moment_scaling}
\ave{x^n}_L \propto L^{D_x(1+n-\tau_x)} + corrections\text{,}
\end{equation}
for $n>\tau_x$, known as \textit{gap scaling}. A moment analysis allows the precise determination of the critical exponents $\tau_x$ and $D_x$, via asymptotic scaling of the moments. The correction terms which scale slower than $L^{D_x(1+n-\tau_x)}$ account for the fact that events of small size $x < x_0$ generally do not follow the \emph{ansatz}
in \Eref{simple_scaling_ansatz}.

The critical exponents for the three avalanche observables considered in this study, avalanche size $s$, avalanche area $a$ and duration time $T$, are extracted using the moment analysis of \Eref{moment_scaling} and reported in \Tref{critical_exponents}. The symbols for the critical exponents follow those that are commonly used in the SOC literature, namely $D\equiv D_s$, $\tau\equiv\tau_s$, $z\equiv D_T$ and $\alpha\equiv\tau_T$. Since the form of correction in \Eref{moment_scaling} is \emph{a priori} unknown, the moments were fitted using the following form with two correction terms, which has proved to work well in a previous analysis \cite{HuynhPruessnerChew:2011}
\begin{equation}
\elabel{moment_fitting}
\langle x^n\rangle(L) = a_1L^{\mu_n} + a_2L^{\mu_n-1} + a_3L^{\mu_n-2}\text{.}
\end{equation}
The fitting produced a goodness-of-fit at least $q=0.195$ for all observables. We extracted scaling exponents $\mu_n$ of the first five moments $n=1,2,\dots,5$ and fitted the resulting values against the linear function $D_x(1+n-\tau_x)$ assuming gap scaling.

% \lecomment{``together with the estimated particle density.'' This hasn't been done. We have the data for the density in both decorrelated Manna Model and the original one for a range of system sizes. The substrate particle density is lower in the decorrelated Manna Model. Letian is able to find the   estimated particle density for the original one. But Letian have not been able to find a suitable form for the correction to scaling for the density in the decorrelated Manna Model.}

The error bar for the exponents extracted in the present work is larger than that reported in one of our earlier studies %\cite{PruessnerHuynh:2012} 
\cite[Tab.~9.3 in][]{Huynh:2013}
because fewer avalanches are used due to the higher demands on CPU-time needed to perform the re-arrangement of particles on the lattice after \emph{every} non-vanishing avalanche. Another reason for the larger error bar is the breakdown of simple scaling to be discussed below. In order to reliably reduce finite size effects, we had to simulate systems up to a size of $L=8193$.

\begin{table}[h]
\small
%\centering
\caption{\tlabel{critical_exponents} Critical exponents for the AMM on a two-dimensional square lattice. All moments were fitted using $\langle x^n\rangle=a_1L^{\mu_n}+a_2L^{\mu_n-1}+a_3L^{\mu_n-2}$ using $n=1,2,\dots,5$ and critical exponent extracted by fitting them against $D_x(1+n-\tau_x)$  . ``$2d$ \textit{twisting}'' refers to the AMM on a two-dimensional lattice closed periodically in one direction and with \textit{twisting} applied as discussed in \sref{initialisation}. ``$2d$ \textit{periodic}'' refers to the AMM with periodic boundary conditions in one direction but without any operations between avalanches. ``$2d$ \textit{original}'' refers to the common AMM on a square lattice open in all directions as studied in \cite{HuynhPruessnerChew:2011}. ``MFT'' refers to mean-field theory results.}
\begin{tabular}{lp{0.2\columnwidth}@{\hskip 0.03\columnwidth}
p{0.2\columnwidth}@{\hskip 0.03\columnwidth}p{0.2\columnwidth}@{\hskip 0.04\columnwidth}l}
\hline\hline
& $2d$ \textit{twisting} & $2d$ \textit{periodic} 
%\cite{PruessnerHuynh:2012} 
%\cite[Tab.~9.3 in][]{Huynh:2013}
\cite{Huynh:2013}
& $2d$ \textit{original} \cite{HuynhPruessnerChew:2011} & MFT \cite{Luebeck:2004} \\
\hline
$D_a$ & $1.799(7)$ & $2.001(1)$ & $1.995(3)$ & $4$ \\
$D$ & $2.028(6)$ & $2.763(8)$ & $2.750(6)$ & $4$ \\
$z$ & $1.050(10)$ & $1.542(5)$ & $1.532(8)$ & $2$ \\
\hline
$\tau_a$ & $1.104(11) $ & $1.378(2)$ & $1.382(3)$ & 3/2 \\ %$\frac{3}{2}$ \\
$\tau$ & $1.017(6)$ & $1.276(2)$ & $1.273(2)$ & 3/2 \\ %$\frac{3}{2}$\\
$\alpha$ & $ 1.138(26) $ & $1.497(8)$ & $1.4896(96)$&  $2$\\
\hline
$-\Sigma_a$ & $ 0.188(19)$ & $0.760(4)$ & $0.76(2)$ \\
$-\Sigma_s$ & $0.035(12)$ & $0.763(8)$ & $0.748(13)$ \\
$-\Sigma_T$ & $0.145(27)$ & $0.766(12)$ & $0.73(4)$ \\
\hline
%$\rho_\infty$ & $0.716045(10)$ & & $0.7170(4)$ \\
\hline\hline
\end{tabular}
\end{table}

% It could be seen that extracted exponents in this set-up of AMM with particles re-arrangements are not the same ones as in the $2d$ \emph{standard} case. The exponents clearly represent a different class that seems to strongly suggest the exact values of $D=2$, $z=1$, $\tau=\tau_a=\alpha=1$, where gas the finite-size scaling exponent of avalanche area $D_a$ appears to suggest a non-trivial value. 
% %The only common value shared by this set-up and the \emph{standard} model is the particle density. 
% It’s fairly to say that AMM with correlations being destroyed is in the different universality class, compared with the $2d$ \emph{original model} and the $2d$ \emph{standard model}.

The exponents can be combined to produce $\Sigma_s=D(1-\tau)$, $\Sigma_a=D_a(1-\tau_a)$ and $\Sigma_T=z(1-\alpha)$, which are identical under the common assumption of a narrow joint distribution of avalanche size, area and duration \cite{JensenChristensenFogedby:1989,ChristensenFogedbyJensen:1991,Luebeck:2000,Pruessner:2012:Book}.

\Tref{critical_exponents} shows a consistent picture, between both the \textit{original} (four open boundaries) and the \textit{periodic} (one open and one periodic direction) Abelian Manna Model as well as within each of them as $\Sigma_a=\Sigma_s=\Sigma_T$. In other words, there is no suggestion that closing one of the directions periodically results in a change of exponents in the \textit{original AMM}. However, the exponents for the \textit{AMM with twisting} clearly deviate. To start with, $D_a$ no longer is close to 2, while $D_a=d$ has been observed for $d\leq4$ \cite{HuynhPruessner:2012b,Bordeu:2019}. In fact, all exponents are much reduced. If our operations to destroy correlations had resulted in an effective random neighbour model, we would expect at least some mean field exponents, which are $D_a=4$, $D=4$, $z=2$, $\tau_a=\frac{3}{2}$, $\tau=\frac{3}{2}$ and $\alpha=2$ \cite{Luebeck:2004}. However, the exponents we found in the \textit{AMM with twisting} deviated from the mean field values more than the \textit{original AMM}. The exponents result in values of $\sum_x$ that are inconsistent with the narrow joint distribution assumption.

To validate our numerics, we compare the estimated first moment to \Eref{first_moment_size} which ought to hold even in the presence of \textit{twisting} because \textit{twisting} does not change the distance of any particle to the open boundaries. This produced perfect agreement. \textit{Twisting} therefore does not have the same effect as bulk-dissipation \cite{BarratVespignaniZapperi:1999}

Similar, consistent results are found for the Abelian Manna Model with \textit{swapping}, \textit{shuffling} and \textit{grinding}, presented in \Aref{appendixa}, although with larger error bars as smaller system sizes are implemented. All the observables studied are consistent within the statistical error under different operations. This is partially in line with our expectations, as all operations destroy correlations while preserving the instantaneous density profile of the substrate. On the other hand, one may expect correlations in the $x-$direction, as destroyed by \textit{twisting} and \textit{grinding}, to have a greater bearing on the scaling behaviour than correlation in the $y-$direction, as destroyed by \textit{swapping}, \textit{shuffling} and \textit{grinding}.

% As mentioned earlier, we also validate our simulation by assessing the value of first moment of avalanche size $\ave{s}$ which can be computed analytically, \Eref{first_moment_size}. In the random walk argument, the random
% walkers perform moves independently, so the site-to-site correlations are
% irrelevant. \Fref{first_moment} shows the consistency between the first moment of the avalanche size and the escape time of a random walker.

% It should also be noted that the exact value for first moment of avalanche size
% $\ave{s}$ as given by \ref{first_moment_size} is not violated because the
% random walk argument still holds.

% \begin{figure}
% \centering
% \includegraphics[width=1\columnwidth]{fig/first_moment.png}
% \caption{\flabel{first_moment}The first moment of avalanche size of AMM with \textit{twisting}.}
% \end{figure}

% LETIAN: Stop editing here. This is all done and dusted. BUT: We need the figures listed below, scollapse_large ... durationcollapse_large
\subsection{Probability density functions}
\Slabel{PDF_analysis}

In addition to a moment analysis, which extracts the critical exponents from the moments of avalanche observables based on the assumption of simple scaling and gap scaling \cite{TebaldiDeMenechStella:1999,Luebeck:2000}, we use the probability density function (PDF) of the observables to qualitatively assess their scaling behaviour in more detail.

% GP started editing here 3 Oct 2023

Normally, the dependence of the PDFs $\mathcal{P}_x(x,L)$ on the system size $L$ is accounted for by the \emph{single} characteristic scale $L^{D_x}$ entering into the scaling function $\mathcal{G}_{x}(x/L^{D_x})$, \Eref{simple_scaling_ansatz}. This function can normally be revealed in a data collapse \cite{Pruessner:2012:Book} by plotting $\mathcal{P}_x(x,L)x^{\tau_x}$ against  $x/L^{D_x}$, which shows $\mathcal{G}_{x}(x/L^{D_x})$ provided $x\gg x_0$, the lower cutoff.

% scollapse_large is the figure that shows the collapse on the basis of the exponents extracted from moments. It also shows \smin, \scrosslow, \scrossup and \smax (or whatever can sensibly be shown). It shows as a dashed line the apparent power law investigated in scollapse_small

% scaling_smax is the figure that shows the scaling of smin, smax, scrosslow and scrossup

% scollapse_small is the figure that shows the PDF with the y axis rescaled in the main panel and both axes rescaled in the inset. It is the current figure 6a. Again, should show as much as possible of smin, smax, scrosslow and scrossup

% areacollapse_large is a collapse of the area PDF on the basis of the exponents from moments, similar to scollapse_large but without the smin, smax etc marker.  But indicate the apparent power law (pretty much like 4a, but with both axis rescaled for a full collapse).

% durationcollapse_large is a collapse of the duration PDF on the basis of the exponents from moments, similar to scollapse_large but without the smin, smax etc marker. But indicate the apparent power law (pretty much like 4c, but with both axis rescaled for a full collapse).

\newcommand{\smin}{s_{\text{min}}}
\newcommand{\smax}{s_{\text{max}}}
\newcommand{\scrosslow}{s_{\times<}}
\newcommand{\scrossup}{s_{\times>}}

\begin{figure}
% \centering
\includegraphics[width=1\columnwidth]{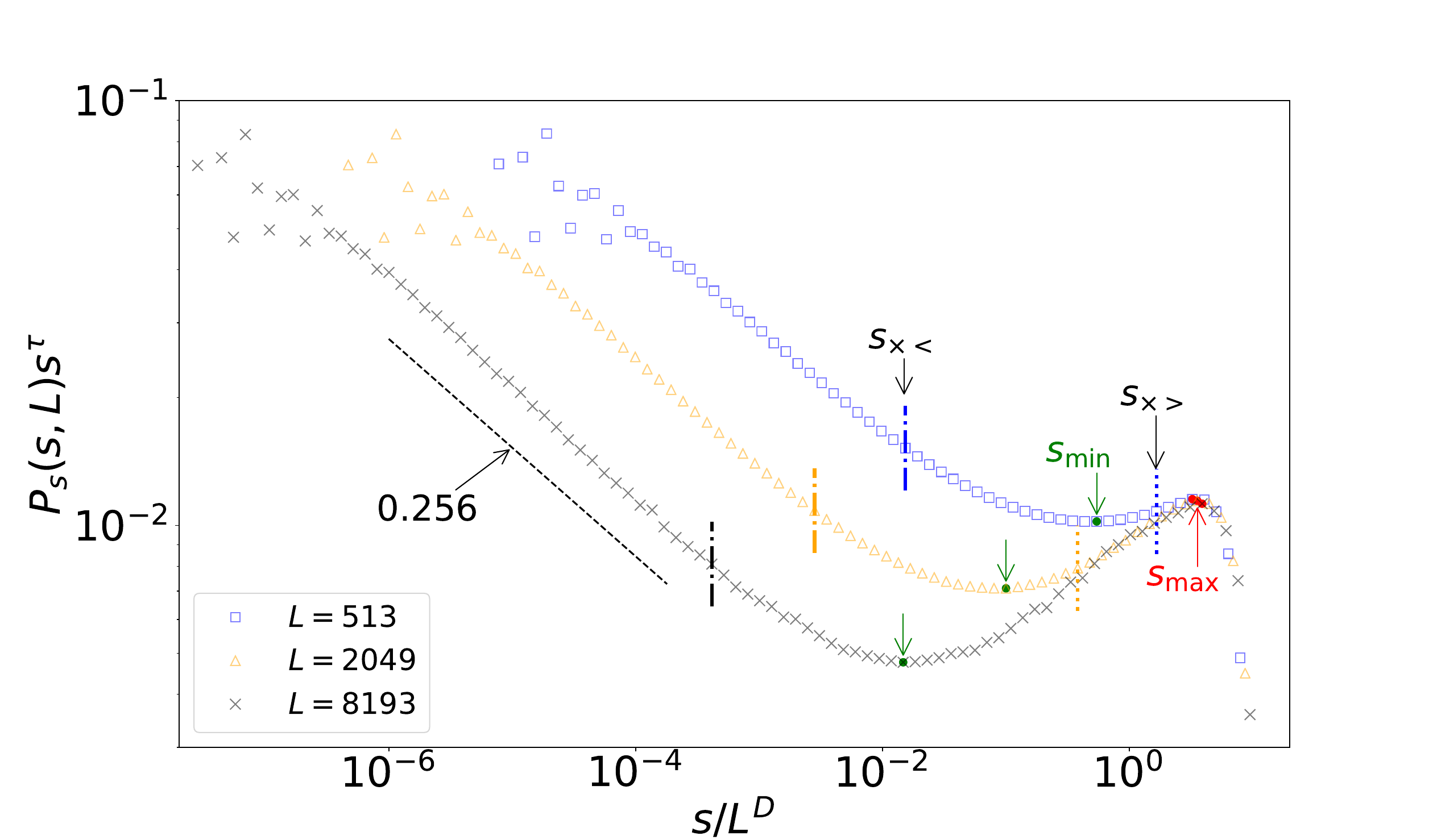}
\caption{\flabel{scollapse_large} Attempted of a data collapse with $\tau=1.017$ and $D=2.028$, \Tref{critical_exponents}. Minima are marked by green arrows pointing downwards. Maxima (collapsed together) are marked by the red arrow pointing upwards. Upper and lower crossovers  $\scrossup$ and $\scrosslow$ are marked by dotted and dash-dotted lines respectively. The dashed line indicates an apparent exponent of $1.017+0.256=1.273$, the exponent reported for $\tau$ in the \textit{original AMM}.  }
\end{figure}

% \begin{figure}[!htb]
% %It's annoying that the subcaptions are automatically centre aligned.
% \subfigure[\flabel{ffm_head} The rescaled histogram of the avalanche size $\mathcal{P}_{s}(s;L)s^{\tau}$, where $\tau=1.273$. The minima in different systems collapse while the maxima don't. The inset is the same data but shifting the curves vertically by $L^{0.7}$]{\includegraphics[width=1\columnwidth]
% {fig/fig_6_a.png}}
% % \subfigure[\flabel{ffm_head_shift_x}The rescaled histogram of the avalanche size $\mathcal{P}_{s}(s;L)s^{\tau}$, \vs $s/L^{0.7}$, where $\tau=1.273$. The front part of the PDFs collapse.]{\includegraphics[width=1\columnwidth]
% % {fig/ffm_head_shift_x.png}}
% \subfigure[\flabel{ffm_tail}The rescaled histogram $\mathcal{P}_{s}(s;L)s^{\tau}$ of the avalanche size, where $\tau=1.017$. The maxima in different systems are equally high while the minima don't.The inset is the same data but shifting the curves vertically by $L^{2.028}$ ]{\includegraphics[width=1\columnwidth]
% {fig/fig_6_b.png}}
% % \subfigure[\flabel{ffm_tail_shift_x}The rescaled histogram of the avalanche size $\mathcal{P}_{s}(s;L)s^{\tau}$, \vs $s/L^{2,028}$, where $\tau=1.017$. The rear part of the PDFs collapse.]{\includegraphics[width=1\columnwidth]
% % {fig/ffm_tail_shift_x.png}}
% \caption{\flabel{ffm}The rescaled histograms of the avalanche size.}
% \end{figure}

\Fref{scollapse_large} shows an attempted data collapse of the PDF of the avalanche size. Using the exponents found on the basis of the moments in \Sref{moment_analysis}, \Tref{critical_exponents}, only a partial collapse can be achieved, which covers only the largest avalanches. This well-collapsed region, starting at around $\scrossup(L)$, as shown in \Frefs{scollapse_large} and \ref{fig:scaling_smax}, and terminating at a scale proportional to $\smax$, increases in range on a logarithmic scale, but covers a fraction of events that decreases with $L$, as $\scrossup(L)\propto L$, \Fref{scaling_smax}

The \emph{apparent power law} of the PDF for smaller avalanche sizes is indicated in \Fref{scollapse_large} by a dashed line, which is characterised by an exponent much closer to the original value, \Tref{critical_exponents}. Rescaling the PDF as to collapse this region produces the plot shown in \Fref{scollapse_small}. The collapsing region, delimited by a constant lower cutoff $s_0$ and the lower crossover $\scrosslow(L)$ covers an increasing fraction of events, but never the largest. As can be seen in \Fref{scaling_smax}, the lower crossover $\scrosslow(L)$ scales roughly like the local minimum $\smin$ shown in \Fref{scollapse_large}.

The PDF of the avalanche size thus exhibits two different scaling regimes where the scaling for small and intermediate avalanches follows that with the steeper exponents of the \textit{original AMM} (\Tref{critical_exponents}), while larger avalanches follow a flatter distribution with the smaller exponents reported for the \textit{AMM with twisting} (\Tref{critical_exponents}). In other words, simple scaling is broken. 

\begin{figure}
% \centering
\includegraphics[width=1\columnwidth]{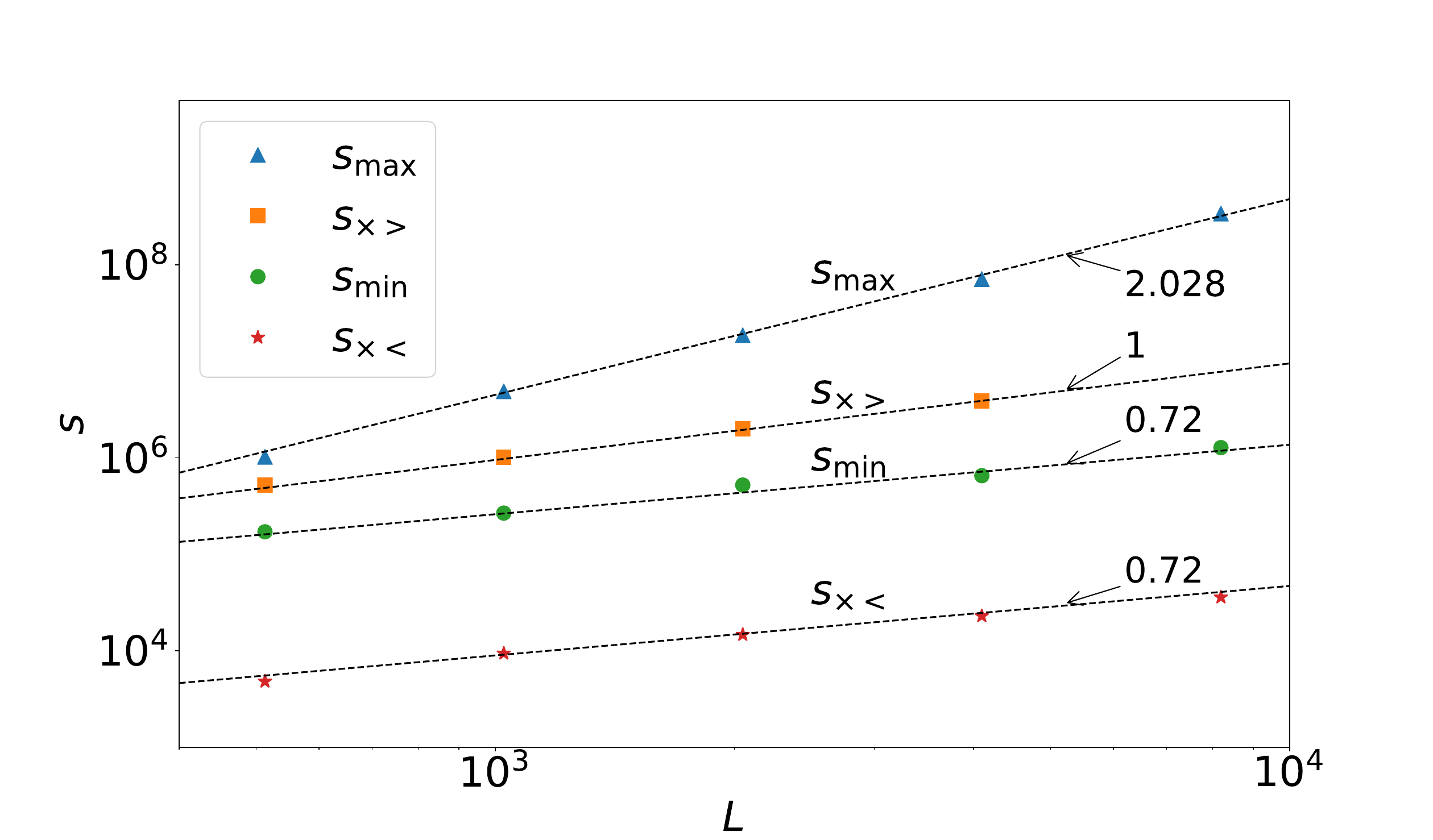}
\caption{\flabel{scaling_smax} The scaling of $\smax$, $\scrossup$, $\smin$ and $\scrosslow$.}
\end{figure}

In total we identify four scales in addition to $s_0$, namely $\scrosslow$, $\smin$, $\scrossup$ and $\smax$, which according to \Fref{scaling_smax} scale roughly like power laws of $L$. While $s_0$ is constant, $\scrosslow$ scales approximately like $L^{0.72}$ and is proportional to $\smin$. This is the region governed by the steeper powerlaw of the \textit{original AMM}. The region from $\scrossup\propto L$ to a scale proportional to $\smax\propto L^{2.028}$ is governed by the shallower power law extracted by the moment analysis \Sref{moment_analysis} . 
At very large avalanche sizes, the PDF is shaped by the scaling function $\GC(s/s_c)$ with upper cutoff $s_c\propto\smax$.
In summary, the PDF of the avalanche size behaves approximately like
\begin{multline}
\elabel{multi_scaling}
\mathcal{P}_{s}(s;L)  \\
=\begin{cases} 
f(s;L) &  s < s_{0} \\
s^{-1.273}  & s_{0}< s < \scrosslow\propto\smin\propto L^{0.72} \\
%s^{-1.017} \mathcal{G}_{s}(s/\smax)  & \scrossup\propto L < s 
s^{-1.017} \mathcal{G}_{s}(s/s_c)  & \scrossup\propto L < s\\  
& \text{with }\ s_c\propto\smax\propto L^{2.028}
\end{cases}
\end{multline}

% \begin{multline}
% \elabel{multi_scaling}
% \mathcal{P}_{s}(s;L)  \\
% =\begin{cases} 
% f(s;L) &  s < s_{0} \\
% s^{-1.273}  & s_{0}< s < \scrosslow\propto\smin\propto L^{0.72} \\
% s^{-1.017} \mathcal{G}_{s}(s/\smax)  & \scrossup\propto L < s < \smax\propto L^{2.028} 
% \end{cases}
% \end{multline}
\begin{figure}
% \centering
\includegraphics[width=1\columnwidth]{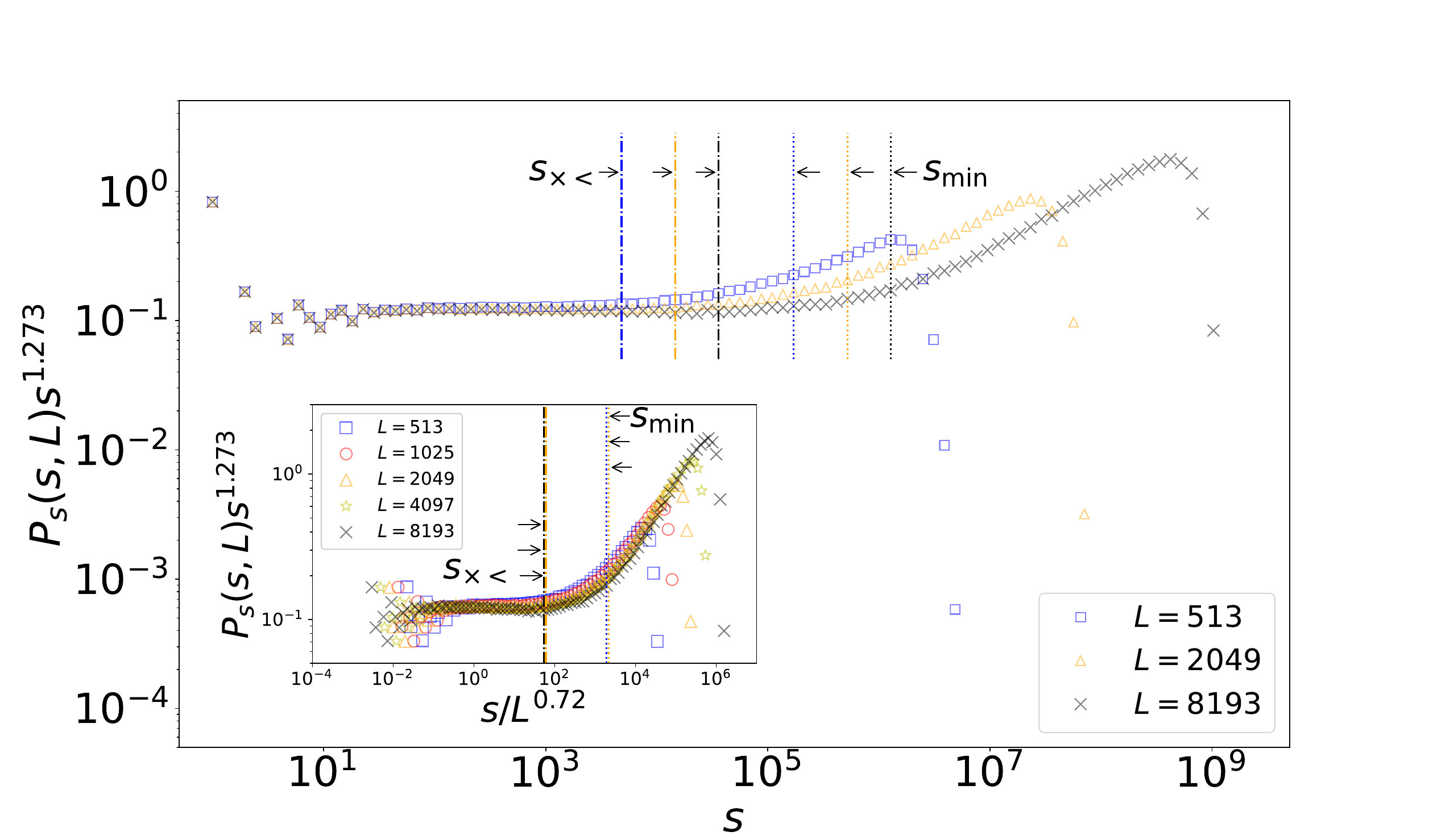}
\caption{\flabel{scollapse_small} The rescaled histogram of the avalanche size $\mathcal{P}_{s}(s;L)s^{\tau}$, where $\tau=1.273$. $\smin$ and $\scrosslow$ are marked by dotted and dash-dotted lines respectively. The inset is the same data but shifting the curves vertically by $L^{0.72}$.     }
\end{figure}
\Fref{scollapse_large} and the inset of \Fref{scollapse_small} demonstrate that the data in different regions can individually be collapsed according to \Eref{multi_scaling}, but never simultaneously for all avalanche sizes. The majority of events have a size less than $\scrosslow$, where the scaling exponent $\tau$ is close to that of the \textit{original AMM}. In other words the majority of events is not affected by the destruction of correlations due to \textit{twisting}. The moments measured in \Sref{moment_analysis} on the other hand are dominated by the scaling of the largest events for any moment $n>\tau-1$, \cite{Pruessner:2012:Book}, and therefore ``report" the exponents characterising the PDF in this region. The situation is very similar to that of the Forest Fire Model \cite{JensenPruessner:2002b,Grassberger:2002a,JensenPruessner:2004}, \ie, 
\textit{twisting} breaks simple scaling as it affects 
only the largest avalanches. Other observables, such as the avalanche area and duration, display similar behaviour, \Fref{area_collapse_large} and \Fref{dua_collapse_large}. However, those observables do not display a collapse as clear as that for the avalanche size, \Fref{scollapse_large}.

\begin{figure}[!htb]
\subfigure[\flabel{area_collapse_large} Attempted data collapse of the avalanche area with $\tau_a=1.104$ and $D_a=1.799$, using the exponents from \Tref{critical_exponents}. The line indicates an apparent exponent of $1.104+0.278=1.382$, the exponent reported for $\tau_a$ in the \textit{original AMM}.]
{\includegraphics[width=1\columnwidth]
{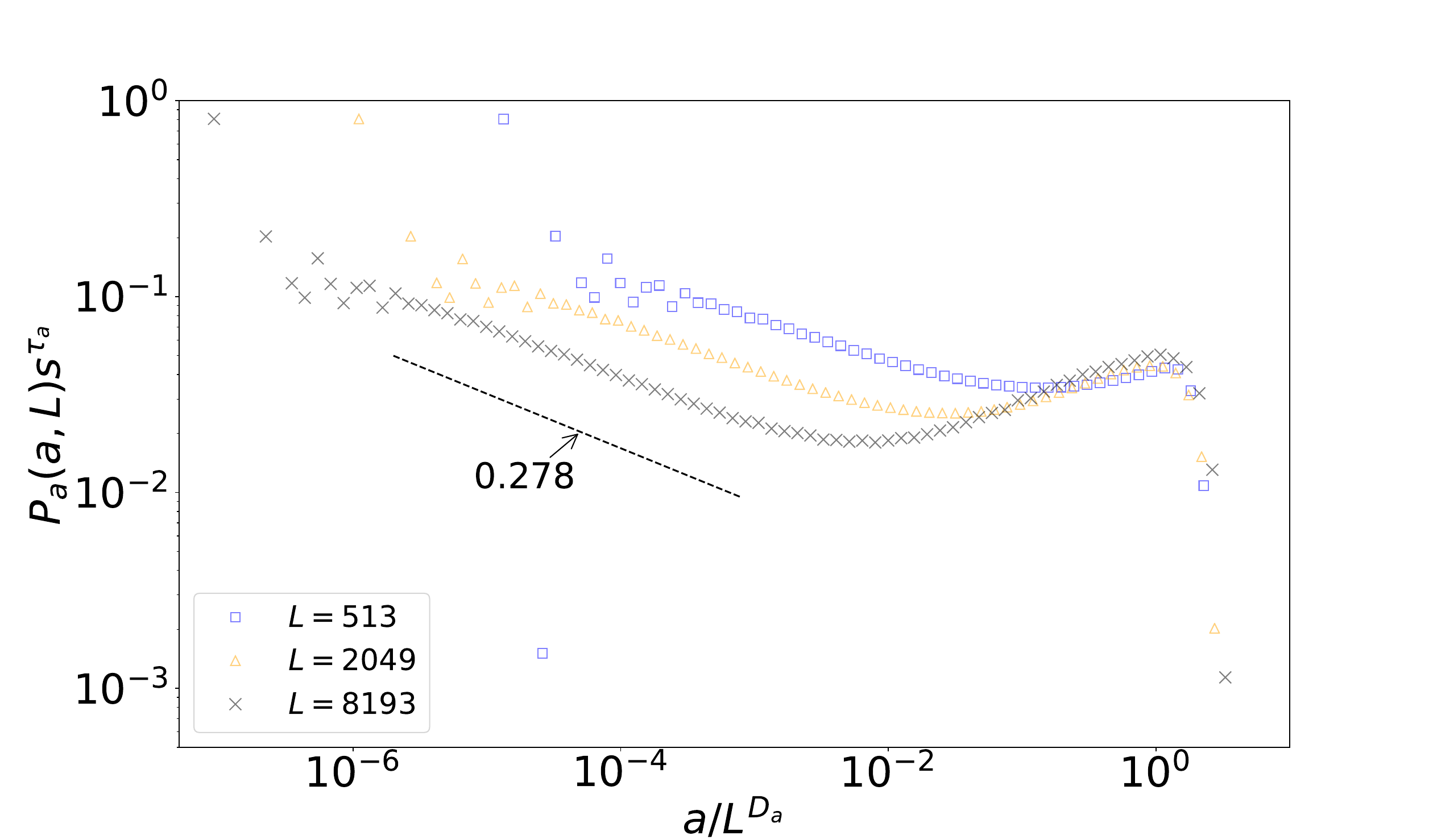}}
\subfigure[\flabel{dua_collapse_large}  Attempted data collapse of avalanche duration with $\alpha=1.138$ and $z=1.050$, using the exponents from \Tref{critical_exponents}. The line indicates an apparent exponent of $1.138+0.3516=1.4896$, the exponent reported for $\alpha$ in the \textit{original AMM}.]
{\includegraphics[width=1\columnwidth]
{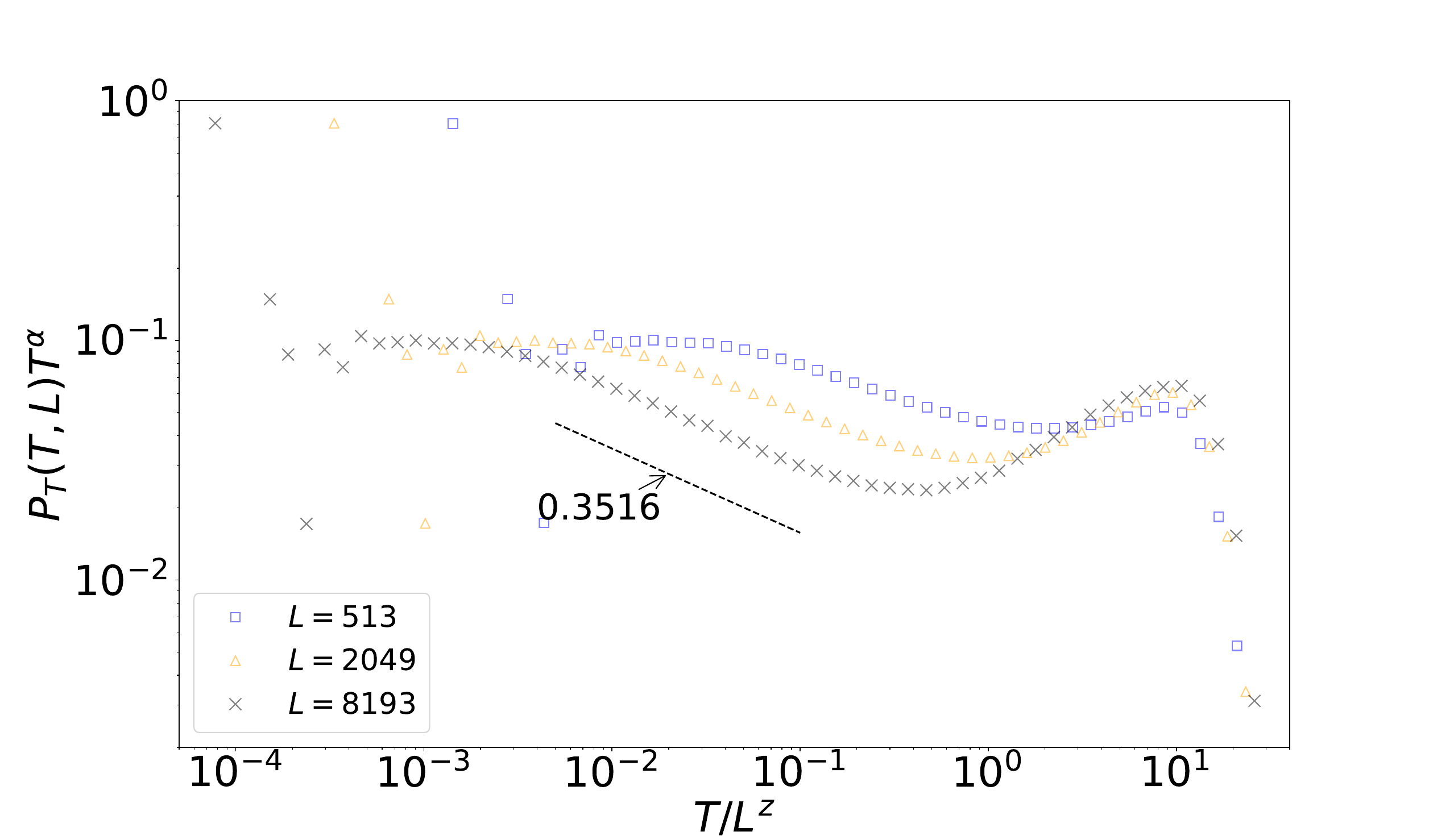}}
\caption{\flabel{pdf_collapses_area_and_duration} Data collapses of avalanche area and avalanche duration .}
\end{figure}

The scaling behaviour suggested in \Eref{multi_scaling} can be used to collapse the data in an exotic way,  \Fref{all_regime_coll}. The ``switch'' from the scaling $\mathcal{P}_{s}(s;L)\propto s^{-1.273}$ up to $\smin \propto L^{0.72}$, to  $\mathcal{P}_{s}(s;L)\propto s^{-1.017} \mathcal{G}_{s}(s/L^{2.028})$ from around $\scrossup\propto L$ is implemented by the exponential $\exp{(-A\cdot s/L)}$ where $A$ is a constant chosen to be $0.005$ here. We have experimented with other sigmoid functions, which produce similar results. In particular, the apparent non-monotonicity of the $x$-axis can be cured for the range of $s$ and $L$ considered.

\begin{figure}
% \centering
\includegraphics[width=1\columnwidth]{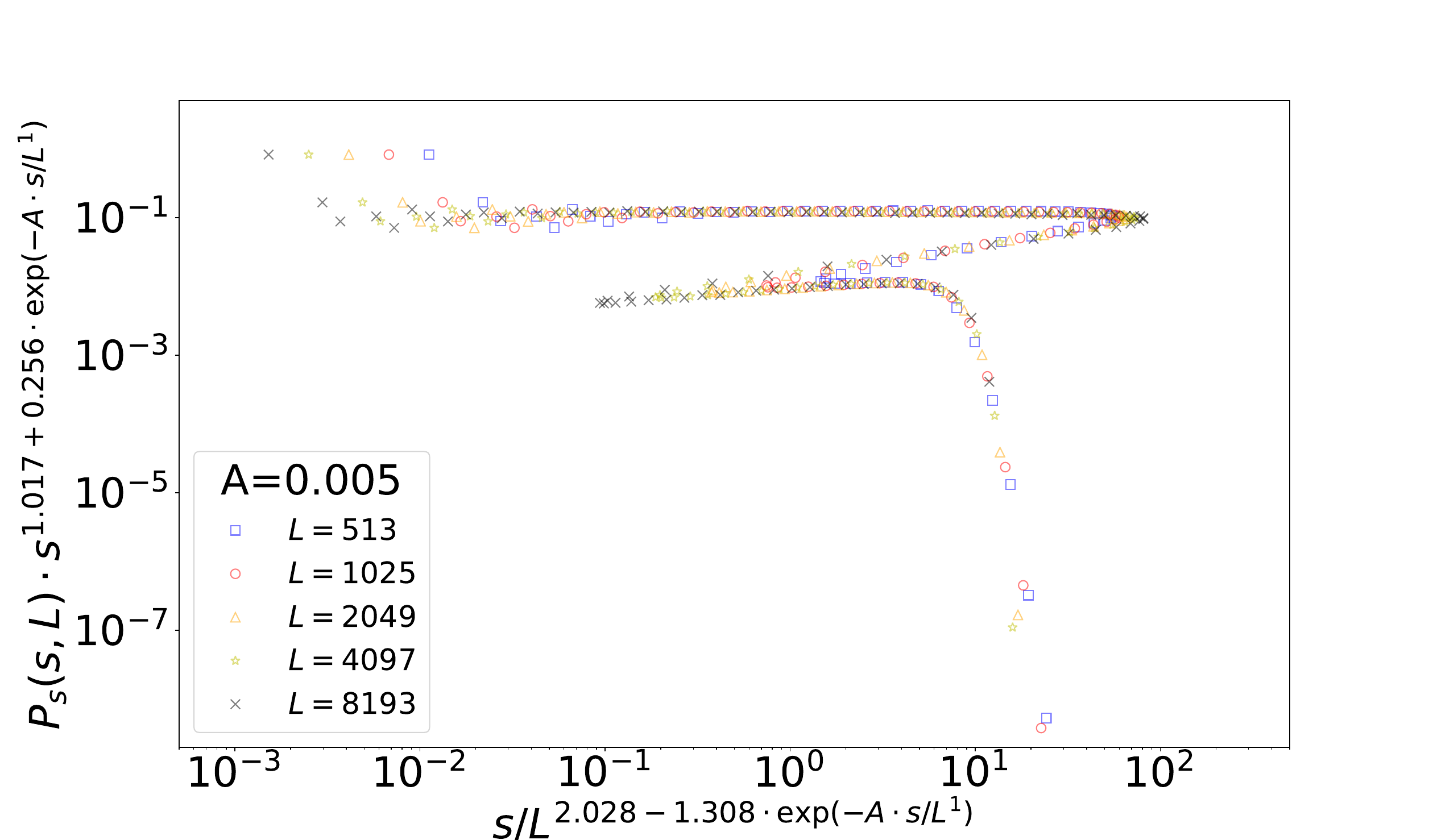}
\caption{\flabel{all_regime_coll} Data collapse using an exponential function which triggers the ``switch'' from the scaling for small avalanches to that of larger avalanches.}
\end{figure}

\section{DISCUSSION}
\Slabel{discussion}
%\subsection{Validity of moment analysis}
\subsection{Different exponents}
\Slabel{validity}

The moment analysis of \Sref{moment_analysis} (\Tref{critical_exponents}) 
produce exponents for the \textit{AMM with twisting} that are different from those of the \textit{periodic AMM} and the \textit{original AMM}. All exponents, even $D_a$, which is generally expected to be identical to the spatial dimension of the lattice \cite{Luebeck:2004,Pruessner:2012:Book} are observed to be significantly smaller in the model with \textit{twisting}.
The other operations to destroy correlations, namely by \textit{swapping}, \textit{grinding} and \textit{shuffling}, as described in \Sref{initialisation}, produce moments whose exponents are consistent with the results for \textit{twisting}, \Tref{critical_exponents_all}. This indicates that there is universality underpinning a ``\textit{decorrelated AMM}".

On the other hand, in \Sref{PDF_analysis} it was demonstrated that the PDF of the avalanche size and, similarly, for the avalanche duration and area, shows two distinct scaling regimes, as was summarised in \Eref{multi_scaling}. This is because the moment analysis is sensitive only to the scaling of the largest events and therefore reports only the scaling exponents in this region. The situation is similar to that in the Drossel-Schwabl Forest Fire Model, which shows systematic,``clean" scaling exponents in a moment analysis and yet distinct scaling regimes in the PDF \cite{JensenPruessner:2004}. What makes the Forest Fire Model different from the present \textit{AMM with twisting} is the remarkably clean collapse of the PDF of the largest avalanche sizes in the latter, \Fref{scollapse_large}, compared to the rather poor collapse of the PDF in the former \cite{JensenPruessner:2004} (Fig.~17).

% \gcomment{Letian, check that we always discuss and show things in the same order: Size, area, duration.} \lecomment{Done.}

As shown in \Sref{PDF_analysis}, \Frefs{scollapse_large} and \ref{fig:pdf_collapses_area_and_duration}, 
the apparent exponents \cite{ChristensenETAL:2008} characterising the PDFs of the size, area and duration of small avalanches, $\tau$, $\tau_a$ and $\alpha$ respectively, are consistent with those found in the \textit{original model}, \Tref{critical_exponents}, albeit less clearly so for area and duration. The majority of events occurs in this small and intermediate regime.

We thus arrive at the following picture: The \textit{AMM with twisting} has two different scaling regimes. Small avalanches display a scaling, characterised by the exponents $\tau$, $\tau_a$ and $\alpha$, that are consistent with the \textit{original model}. Large avalanches in the model with \textit{twisting} display a scaling with modified exponents. Not only the exponents $\tau$, $\tau_a$ and $\alpha$ that normally determine the slope of the PDF are affected, but also the cutoff exponents, $D$, $D_a$ and $z$. As far as the regime of large avalanches is concerned, all exponents are clearly reduced, \Tref{critical_exponents}. The values are inconsistent with the assumption of a narrow joint distribution, \Sref{moment_analysis}, as is indicated by the different values of $\Sigma_x$ in \Tref{critical_exponents}.

\subsection{Role of correlations}
\Slabel{role_correlations}

Attributing the results above to the destruction of correlations by the \textit{twisting}, small avalanches, and thus the majority of events, appear to be much less affected by the ``decorrelation" than large avalanches. This is a surprising outcome, as one may expect that large avalanches are not affected by a change in the spatial correlations, based on the following arguments: Firstly, the spatial correlations in the substrate are rather weak. In \cite{WillisPruessner}, the amplitude of the correlations is found to be convergent in the limit of large system size, implying that the correlations do not increase in larger systems. Secondly, during a large avalanche, the substrate is rearranged many times, as many sites topple an increasing number of times during avalanches, which can be seen by $s\ge a$ generally and similarly for the cutoffs of size and area, $L^{D} \gg L^{D_a}$ in large $L$, so that the characteristic number of topplings per site taking part diverges with the system size and in fact scales like $L^{D-D_a}$ for the largest avalanches, with $D-D_a=0.762(9)$ for the \textit{periodic model} and $D-D_a=0.229(13)$ for the \textit{AMM with twisting} according to \Tref{critical_exponents}. This suggests the characteristics of large avalanches are generally determined by the dynamics, not by the substrate. 

\begin{figure}
% \centering
\includegraphics[width=1\columnwidth]{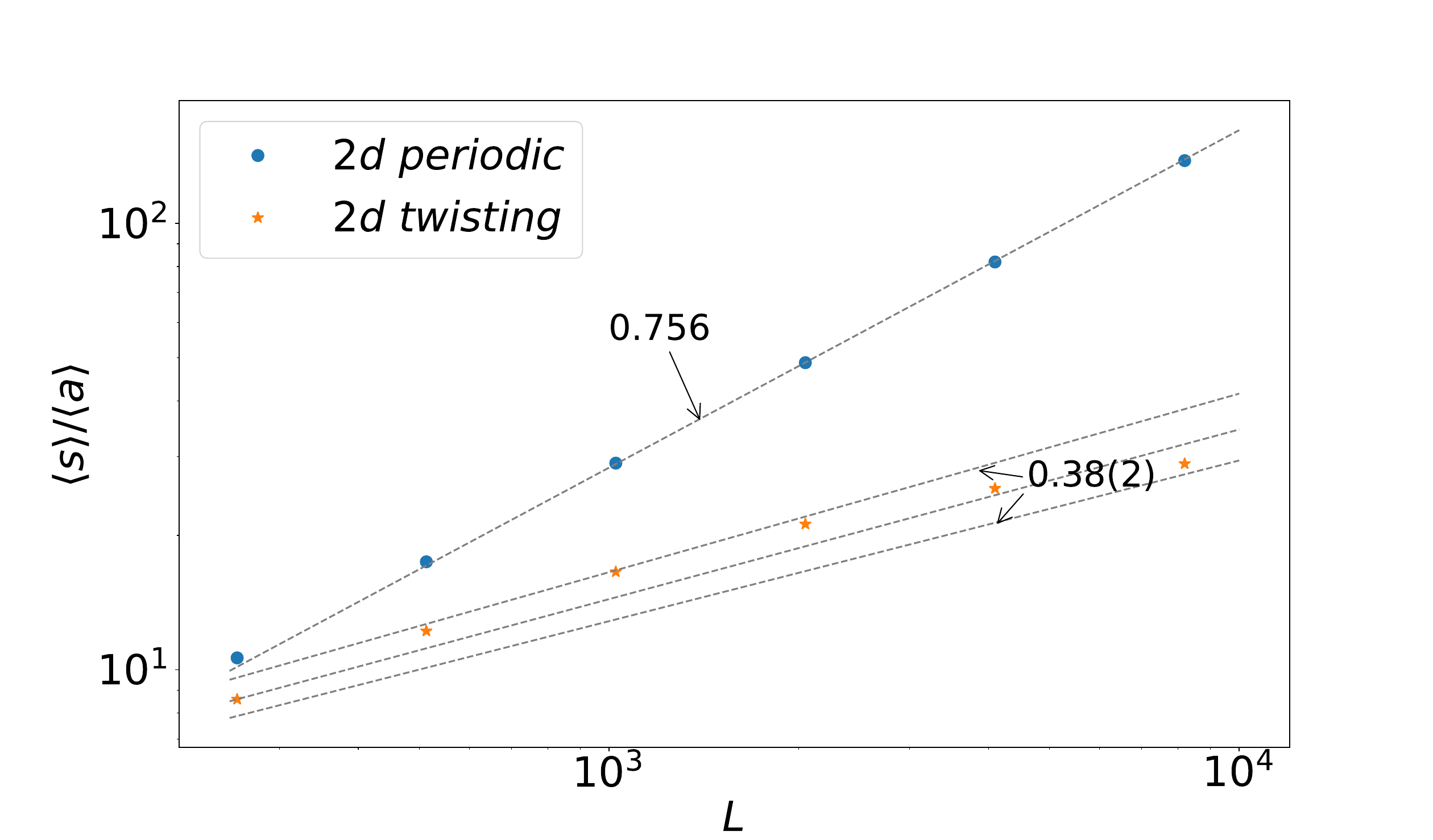}
\caption{\flabel{s_a_tw} Estimate of the average number of topplings, $\langle s \rangle/\langle a \rangle$, for a range of system sizes. This ratio is always larger in the \emph{periodic model}, compared to the AMM with \emph{twisting}. The exponents indicated are $D(2-\tau)-D_a(2-\tau_a)$, which are $0.756(9)$ for the \emph{periodic model} and $0.38(2)$ for the \emph{AMM with twisting} based on \Tref{critical_exponents}, as $\ave{s}\propto L^{D(2-\tau)}$ and $\ave{a}\propto L^{D_a(2-\tau_a)}$.}
\end{figure}
% One argument might be that after an avalanche, an occupied site is more likely to have an empty neighbor. This can be seen in the correlation diagram (Fig.~2 in Ref.~\cite{WillisPruessner}) that correlation between neighbors has a weak but \emph{non-zero} negative value. After the rearrangements of the substrate, the correlations are gone. As a result, an occupied site is more likely to have occupied neighbors, and large avalanches are more likely to occur, compared to the system without the rearrangements. That might be why the statistics of large avalanches are affected by a change in the spatial correlations.

% \gcomment{Letian, add sloped lines in the \Fref{s_a_tw} and confirm my values in main text and caption. }\lecomment{Done.}

The average number of topplings per site across all avalanches is shown in \Fref{s_a_tw} in the form $\ave{s}/\ave{a}$. This ratio should scale with exponent $D(2-\tau)-D_a(2-\tau_a)$, which according to \Tref{critical_exponents} is $0.756(9)$ for the \textit{periodic model} and $0.38(2)$ for the \textit{AMM with twisting}. The small value of the exponent for the latter explains to some extent the clearly visible corrections, that the \textit{AMM with twisting} is suffering from. For the largest system sizes considered, even on average, every site topples tens of times during a single avalanche.

A more conservative estimate of the mean ratio $\ave{s/a}$ gives the scaling $L^{D(2-\tau)-D_a}$ and thus a constant, assuming $a \sim s^{\frac{D_a}{D}}$ based on narrow joint distributions, as well as $\ave{s} \sim L^{2}$ and $D_a = 2$. Given that $s/a$ frequently vanishes, a finite $s/a$ is indicative of large values of $s/a$ for large avalanches.

From the two points above one might expect that in the \emph{periodic AMM} a build-up of correlations in the substrate affects the majority of avalanches from event to event, except for the very large events that create ``their own environment". Destroying these correlations should therefore affect small avalanches more than large ones.

%XXX Maybe discuss s_a_tw here. <s>/<a> goes up, but doesn't look clean for twisting. Maybe that's because of <a>? Because we are saying above that <s> is solid. 

However, this turns out not to be the case. In fact, the opposite happens: Small avalanches seem almost unaffected by the \textit{twisting}, while large avalanches clearly are. Why does this happen? We dismiss the na{\"i}ve explanation that large avalanches are system-spanning, and thus \textit{more} affected by the absence of correlations than small avalanches, as every avalanche takes place on a decorrelated substrate. The fraction of sites that are charged only once or topple only once is generally larger for small avalanches than for large avalanches. Again, the destruction of correlations in the substrate should therefore affect smaller avalanches more than larger ones.

We speculate that a more subtle effect is at work. As can be seen in Fig.~2 in \cite{WillisPruessner}, site-occupation shows weak \textit{anti}-correlations, which means that occupied sites effectively \textit{repel} occupied sites. As \textit{twisting} destroys these anti-correlations, activity spreads more easily to occupied nearest neighbours so that large avalanches are triggered prematurely. This is corroborated by the significant reduction of the exponent $D$ characterising the cutoff of the avalanche size distribution. Avalanches in the \textit{AMM with twisting} are uncharacteristically small. In fact $D$ barely exceeds $d=2$, although it still exceeds $D_a$. 

% GP: I think the next point is interesting, but difficult to convey
%Given the anti-correlations, it seems that the destruction of correlations does not generally result in large occupied patches in the \emph{original AMM} being broken up into smaller patches, like a preventative measure to reduce the risk of catastrophically large avalanches. 

%However, it is impossible to distinguish whether the destruction of correlations favours intermediate sized avalanches or suppresses very large ones, as both goes hand-in-hand.

%This insight cannot be based on the first moment of the avalanche size, which is the same with or without twisting. The second moment scaling 

%One might speculate that the destruction of correlations results densely occupied patches in the \emph{original AMM} to break up into smaller patches, such that avalanches are triggered more often, but are generally smaller.

% \gcomment{Letian, give me an estimate for the substrate density in the AMM with twisting and the original model with L=513, 2047 and 8193. Also calculate the standard deviation of the density (not the error, let's discuss, this is $\sqrt{\ave{\zeta}^2-\ave{\zeta}^2}$}. \lecomment{Done.}

The largest avalanches in the decorrelated \textit{AMM with twisting} are thus small compared to those in the \textit{original model.} Those very large ones, that take place in the \textit{original model}, are effectively suppressed in the \textit{AMM with twisting} because the system is more frequently flushed of fuel and we end up exploring the Manna Model in a regime that would otherwise be considered as sub-critical. The resulting PDF may thus be considered as the PDF of the \textit{periodic AMM} cut off at around $\scrosslow$ and characterised in the tail by an interplay of the residual correlations and the dynamics, which produces avalanches that are barely system spanning, $D_a<2$ and $D$ barely exceeding $2$.

% This is further supported by the substrate particle density (\Tref{critical_exponents}) being lower in the decorrelated Manna Model compared to the original one. \lecomment{This hasn't been done. We have the data for the density in both decorrelated Manna Model and the original one for a range of system sizes. The substrate particle density is lower in the decorrelated Manna Model. But Letian have not been able to find a suitable form for the correction to scaling for the density in the decorrelated Manna Model.}

% \gcomment{We really should have only one term for the models. AMM with twisting, original AMM and standard AMM? All in italics? If we keep using other terms, the reader will think these are different models. Letian, please stream-line!} \lecomment{Done.}

\begin{figure}
% \centering
\includegraphics[width=1\columnwidth]{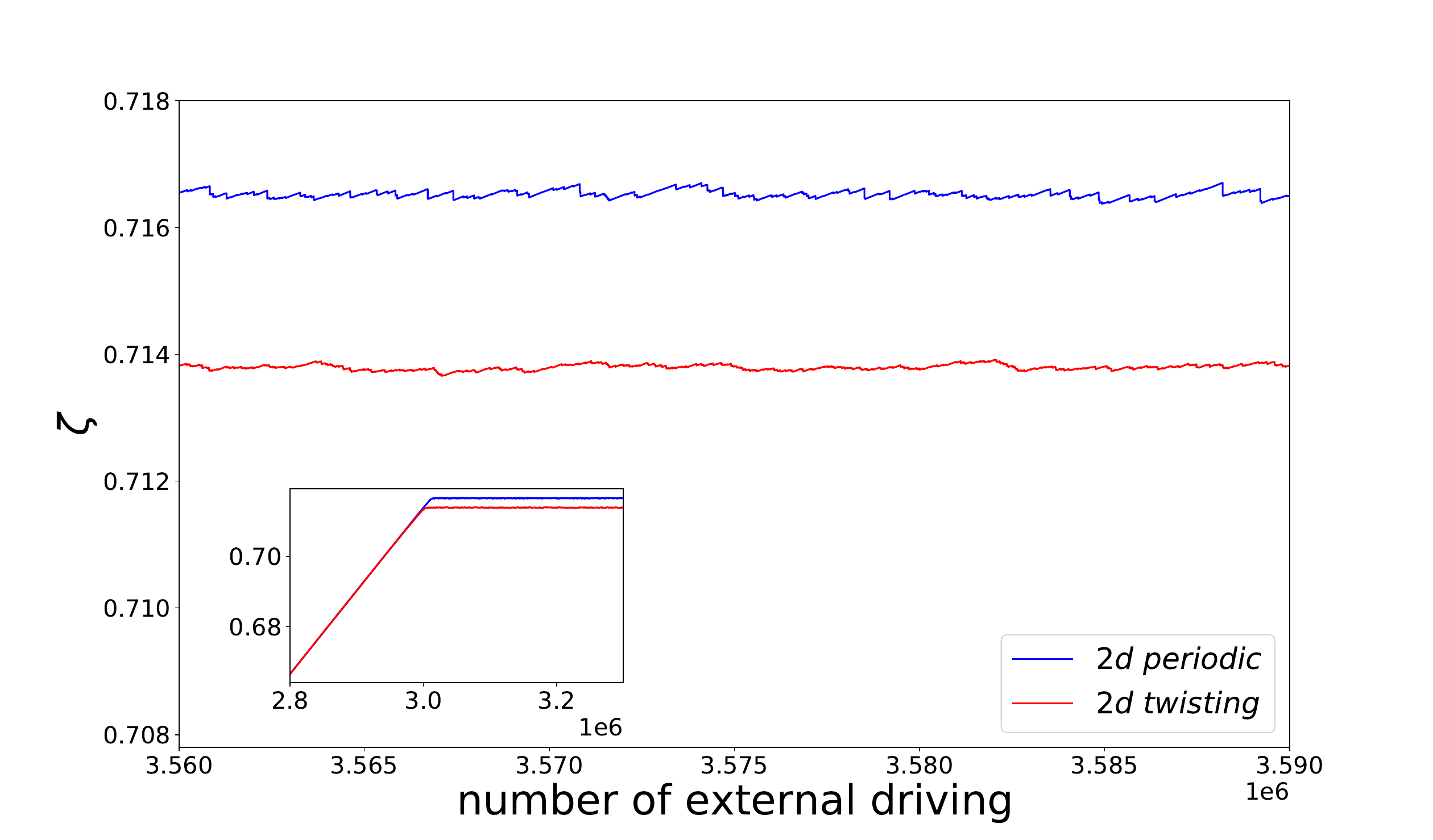}
\caption{\flabel{eva_density} The instantaneous particle density as a function of the number of external driving in a $2049\times2049$ system. The inset shows the same data but on a lager scale of the number of external drivings. }
\end{figure}

To illustrate this more vividly, we show the instantaneous particle density in a single realisation as a function of the number of external drivings in a $2049\times2049$ system in \Fref{eva_density}. As shown in the inset, the instantaneous particle density in the \textit{AMM with twisting} follows that of the \textit{periodic model} but saturates earlier. As can be seen in the main panel, the substrate density of $\zeta=0.71383(6)$ 
% \gcomment{Letian, need error bar!} \lecomment{Done.} 
in the \textit{AMM with twisting} is typically much lower than that in the \textit{periodic model}, $\zeta=0.71652(6)$ 
% \gcomment{Letian, need error bar!} \lecomment{Done.}
, amounting to about $11294$ particles ``missing" in the former. 

% \gcomment{Letian, please confirm std of density.} \lecomment{Done.}

The fluctuations of the density are visibly stronger in the \textit{periodic model} compared to the \textit{AMM with twisting}, $\sigma(\zeta)=6.07\cdot 10^{-5}$ versus $\sigma(\zeta)=5.69\cdot 10^{-5}$. Yet, they are orders of magnitude smaller than the difference between the two densities of $2.69\cdot 10^{-3}$. The 
\textit{periodic model} also has more rugged density fluctuations which display sharp, sudden drops as the very largest of avalanches occur, ejecting many particles from the system at once. We speculate that the \textit{AMM with twisting} is thus maintained in a subcritical state, so that reducing the correlations effectively cuts off critical scaling at $\scrosslow$, \Sref{PDF_analysis}.

\section{Conclusion}
\Slabel{conclusion}

In this work, we investigate the role of spatial correlations in the substrate in the Abelian Manna Model (AMM) by destroying them deliberately and observing  the change of the critical behaviour. Several operations,  \textit{twisting}, \textit{swapping}, \textit{shuffling} and \textit{grinding}, are performed in extensive numerical simulations of several different systems sizes. Using moment analysis, we extract critical exponents for different observables, namely avalanche size, area and duration, and show that the destruction of correlations in the substrate results in scaling that differs from the \textit{periodic Abelian Manna Model}. Further assessing the PDFs shows that a crossover takes place from one scaling regime that resembles the \textit{periodic AMM} to another, new one, that is characterised by the exponents extracted from the moment analysis. The existence of two distinct scaling regimes is reminiscent of the Forest Fire Model \cite{JensenPruessner:2002b}. 

Destroying correlations in the Manna Model thus destroys simple scaling. We conclude that substrate correlations are important in the Manna Model, in line with the narrative that the correlated debris left behind by one avalanche shapes the next \cite{WatkinsETAL:2016}, producing non-trivial correlations between and within avalanches \cite{GarciaMillanPruessner:2018}. Any theoretical treatment will thus need to incorporate these correlations.

In future work, one might study the effects of re-arrangements of substrate particles in other, related models. For example, neural networks have been studied intensely from the perspective of self-organised critical phenomena \cite{Hopfield:1994,deArcangelisPerrone-CapanoHerrmann:2006,Herrmann:2016,Jensen:2021}. It has been found experimentally that the connection map in neural network can change over short and long time scales \cite{Finnerty:1999,Sophie:2018,Chklovskii:2004,KoHo:2013}, which effectively destroys the correlations in the substrate and thus strongly affects the system's behaviour.

%Earthquake?

\begin{acknowledgments}
The authors would like to thank Henrik Jensen for interesting discussions, as well as Andy Thomas and Niall Adams for invaluable computing
support. LTC thanks Qing Yao, Connor Robert, Jacob Knight and Thibault Bertrand for useful discussions. HNH acknowledges the support of the A*STAR International Fellowship (2016 - 2018).
\end{acknowledgments}

% \bibliography{amm_corr}% Produces the bibliography via BibTeX.

\clearpage

% XXX GP to continue here
% \newpage
\appendix
\begin{widetext}
\section{Moment Analysis of AMM with \emph{swapping}, \emph{grinding}, and \emph{shuffling}}
\alabel{appendixa}

In this section, we report the results of the \emph{moment analysis} for the AMM with \textit{swapping}, \textit{grinding}, and \textit{shuffling} applied as discussed in \sref{initialisation}. The linear sizes considered for \textit{swapping} and \textit{grinding} operations were $L \in \{257, 513, 1025, 2049, 4097\}$. Due to the higher demands on CPU-time needed to perform the \textit{shuffling} operation, the linear sizes considered for \textit{shuffling} operations were $L \in \{129, 257, 513, 1025, 2049\}$. For this form of destroying the correlations, we were able to fit the critical exponents only for the avalanche size with reasonable goodness-of-fit. The resulting exponent for $\tau$ has an estimated value below unity, which is known to be impossible \cite{ChristensenETAL:2008}, although unity is covered by the error bar.

\begin{table*}[ht]
\caption{\tlabel{critical_exponents_all} 
 Critical exponents and particle density for the AMM under various operations on a two-dimensional square lattice. All moments were fitted using $\langle x^n\rangle=a_1L^{\mu_n}+a_2L^{\mu_n-1}+a_3L^{\mu_n-2}$ using $n=1,2,\dots,5$ and critical exponent extracted by fitting them against $D_x(1+n-\tau_x)$  . ``$2d$ \textit{swapping}'',``$2d$ \textit{grinding}'' and ``$2d$ \textit{shuffling}'' refer to the AMM on a two-dimensional lattice closed periodically in one direction and with decorrelating operation applied as discussed in \sref{initialisation}. ``$2d$ \textit{periodic}'' refers to the AMM with periodic boundary conditions in one direction and  without any operations between avalanches. ``$2d$ \textit{original}'' refers to the AMM on a square lattice open in all directions as studied in \cite{HuynhPruessnerChew:2011}. 
 The goodness-of-fit is at least $q = 0.204$ for all observables considered for \textit{swapping}, at least $q=0.224$ for \textit{grinding} and $q=0.12$ for \textit{shuffling}, in which case exponents were extracted only for the avalanche size distribution. The data for \textit{twisting}, also shown in \Tref{critical_exponents} is shown for comparison. }
 \resizebox{\textwidth}{!}{
\begin{tabular}{l
p{0.2\columnwidth}@{\hskip 0.03\columnwidth}
p{0.2\columnwidth}@{\hskip 0.03\columnwidth}
p{0.2\columnwidth}@{\hskip 0.03\columnwidth}
p{0.2\columnwidth}@{\hskip 0.03\columnwidth}
p{0.2\columnwidth}@{\hskip 0.03\columnwidth}l}
\hline\hline
& $2d$ \emph{swapping}
& $2d$ \emph{grinding}
& $2d$ \emph{shuffling}
& $2d$ \emph{twisting}
& $2d$ \emph{periodic} 
\cite{Huynh:2013}
%\cite[Tab.~9.3 in][]{Huynh:2013}
%\cite{PruessnerHuynh:2012} 
& $2d$ \emph{original} \cite{HuynhPruessnerChew:2011} \\
\hline
$D_a$ & $1.827(48)$ & $1.781(21)$ &        ---  & $1.799(7)$ & $2.001(1)$ & $1.995(3)$ \\
$D$   & $2.034(17)$ & $2.038(15)$ & $2.016(40)$ & $2.028(6)$ & $2.763(8)$ & $2.750(6)$ \\
$z$   & $1.030(32)$ & $1.015(40)$ &        ---  & $1.050(10)$ & $1.542(5)$ & $1.532(8)$ \\
\hline
$\tau_a$ & $1.149(57)$ & $1.096(29)$ &      ---   & $1.104(11)$ & $1.378(2)$ & $1.382(3)$ \\
$\tau$   & $1.021(14)$ & $1.018(13)$ & $0.977(36)$& $1.017(6)$  & $1.276(2)$ & $1.273(2)$ \\
$\alpha$ & $1.100(82)$ & $1.08(10)$&      ---   & $1.138(26)$ & $1.497(8)$ & $1.4896(96)$ \\
\hline
$-\Sigma_a$ & $0.273(105)$ & $0.171(51)$  &         ---  & $0.188(19)$ & $0.760(4)$ & $0.76(2)$ \\
$-\Sigma_s$ &  $0.042(28)$ & $0.036(27)$  & $-0.047(72)$ & $0.035(12)$ & $0.763(8)$ & $0.748(13)$ \\
$-\Sigma_T$ & $ 0.103(84)$ & $0.08(10)$ &         ---  & $0.145(27)$ & $0.766(12)$ & $0.73(4)$ \\
\hline
%$\rho_\infty$ & $0.716045(10)$ & & $0.7170(4)$ \\
\hline\hline
\end{tabular}
}
\end{table*}

\clearpage
\end{widetext}
\end{document}

%% file: macrosGunnar.tex
\newcommand{\newevenside}{
        \ifthenelse{\isodd{\thepage}}{\newpage}{
        \newpage
        \phantom{placeholder} % doesn't appear on page
        \thispagestyle{empty} % if no header/footer wanted
        \newpage
        }}

\newcommand{\ie}{{\it i.e.~}} % I prefer traditional way of using \it command

\newcommand{\etal}{{\it et al.}}

\newcommand{\Cref}[1]{Chapter~\ref{chap:#1}}

\newcommand{\Slabel}[1]{\label{sec:#1}} % no idea why \slabel cannot be used
\newcommand{\sref}[1]{Sec.~\ref{sec:#1}}
\newcommand{\Sref}[1]{Sec.~\ref{sec:#1}}

\newcommand{\alabel}[1]{\label{sec:#1}} % for appendix
\newcommand{\Aref}[1]{Appendix~\ref{sec:#1}}

\newcommand{\elabel}[1]{\label{eq:#1}}

\newcommand{\Eref}[1]{Eq.~(\ref{eq:#1})}

\newcommand{\flabel}[1]{\label{fig:#1}}

\newcommand{\Fref}[1]{Fig.~\ref{fig:#1}}

\newcommand{\tlabel}[1]{\label{table:#1}}

\newcommand{\Tref}[1]{Table~\ref{table:#1}}

\newcommand{\ave}[1]{\left\langle #1 \right\rangle}

\newcommand{\GC}{\mathcal{G}}